\documentclass[12pt]{article}

\usepackage{graphics,epsfig}

\newcommand{\bea}{\begin{eqnarray}}
\newcommand{\eea}{\end{eqnarray}}
\newcommand{\beq}{\begin{equation}}
\newcommand{\eeq}{\end{equation}}
\newcommand{\nn}{\nonumber}
\newcommand{\gev}{{\rm GeV}}
\newcommand{\mev}{{\rm MeV}}
\newcommand{\msb}{\overline{\rm{MS}}}

\def\dfrac#1#2{{\displaystyle {#1 \over #2}}}

\def\simge{\mathrel{\rlap{\raise 0.511ex \hbox{$>$}}{\lower 0.511ex
 \hbox{$\sim$}}}}
\def\simle{\mathrel{\rlap{\raise 0.511ex \hbox{$<$}}{\lower 0.511ex
 \hbox{$\sim$}}}}
\def\slash#1{\setbox0=\hbox{$#1$}\dimen0=\wd0 \setbox1=\hbox{/} \dimen1=\wd1
 \ifdim\dimen0>\dimen1 \rlap{\hbox to \dimen0{\hfil/\hfil}} #1
 \else \rlap{\hbox to \dimen1{\hfil$#1$\hfil}} / \fi}

\def\nic{a}
\def\ors{b}
\def\rmii{c}
\def\mns{d}
\def\val{e}
\def\rmiii{f}
\def\liv{g}
\def\cern{h}
\def\mun{i}
\def\zur{j}

\begin{document}

\begin{titlepage}
{\vspace{-0.5cm} \normalsize
\hfill \parbox{90mm}{DESY 07-148, FTUV-07-2709, IFIC/07-57, \\
                     MS-TP-07-23, RM3-TH/07-11, ROM2F/2007/16,\\
                     SFB/CPP-07-58, TUM-HEP-676/07}}\\[10mm]
\begin{center}
  \begin{LARGE}
    \textbf{Light quark masses and pseudoscalar decay\\constants from $N_f=2$
            Lattice QCD\\with twisted mass fermions} \\
  \end{LARGE}
\end{center}

\vskip 0.5cm
\begin{figure}[h]
  \begin{center}
    \includegraphics[draft=false]{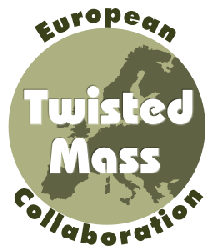}
  \end{center}
\end{figure}

\vspace{-0.8cm}
\baselineskip 20pt plus 2pt minus 2pt
\begin{center}
  \textbf{
    B.~Blossier$^{(\nic)}$,
    Ph.~Boucaud$^{(\ors)}$,
    P.~Dimopoulos$^{(\rmii)}$,
    F.~Farchioni$^{(\mns)}$,
    R.~Frezzotti$^{(\rmii)}$,
    V.~Gimenez$^{(\val)}$,
    G.~Herdoiza$^{(\rmii)}$,
    K.~Jansen$^{(\nic)}$,
    V.~Lubicz$^{(\rmiii)}$,
    C.~Michael$^{(\liv)}$,
    D.~Palao$^{(\val)}$,
    M.~Papinutto$^{(\cern)}$,
    A.~Shindler$^{(\nic)}$,
    S.~Simula$^{(\rmiii)}$,
    C.~Tarantino$^{(\mun)}$,
    C.~Urbach$^{(\liv)}$,
    U.~Wenger$^{(\zur)}$}
\\
\end{center}

\begin{center}
  \begin{footnotesize}
    \noindent 

$^{(\nic)}$ NIC, DESY, Zeuthen, Platanenallee 6, D-15738 Zeuthen, Germany
\vspace{0.2cm}

$^{(\ors)}$ Laboratoire de Physique Th\'eorique (B\^at.~210), Universit\'e de
Paris XI,\\ Centre d'Orsay, 91405 Orsay-Cedex, France
\vspace{0.2cm}

$^{(\rmii)}$ Dip. di Fisica, Universit{\`a} di Roma Tor Vergata and INFN, Sez.
di Roma Tor Vergata,\\ Via della Ricerca Scientifica, I-00133 Roma, Italy
\vspace{0.2cm}

$^{(\mns)}$ Universit\"at M\"unster, Institut f\"ur Theoretische Physik,
\\Wilhelm-Klemm-Strasse 9, D-48149 M\"unster, Germany
\vspace{0.2cm}

$^{(\val)}$ Dep. de F\'{\i}sica Te\`{o}rica and IFIC, Univ. de Val\`{e}ncia,\\
Dr.Moliner 50, E-46100 Burjassot, Spain
\vspace{0.2cm}

$^{(\rmiii)}$ Dip. di Fisica, Universit{\`a} di Roma Tre and INFN, Sez. di Roma
III,\\ Via della Vasca Navale 84, I-00146 Roma, Italy
\vspace{0.2cm}

$^{(\liv)}$ Theoretical Physics Division, Dept. of Mathematical Sciences,
\\University of Liverpool, Liverpool L69 7ZL, UK
\vspace{0.2cm}

$^{(\cern)}$ CERN, Physics Department, Theory Division, CH-1211 Geneva 23,
Switzerland
\vspace{0.2cm}

$^{(\mun)}$ Physik Department, Technische Universit\"at M\"unchen, D-85748
Garching, Germany
\vspace{0.2cm}

$^{(\zur)}$ Institute for Theoretical Physics, ETH Zurich, CH-8093 Zurich,
Switzerland

    \end{footnotesize}
  \end{center}

\end{titlepage}

\begin{abstract}{
We present the results of a lattice QCD calculation of the average up-down and
strange quark masses and of the light meson pseudoscalar decay constants with
$N_f=2$ dynamical fermions. The simulation is carried out at a single value of
the lattice spacing with the twisted mass fermionic action at maximal twist,
which guarantees automatic ${\cal O}(a)$-improvement of the physical quantities.
Quark masses are renormalized by implementing the non perturbative RI-MOM
renormalization procedure. Our results for the light quark masses are
$m_{ud}^{\msb}(2\ \gev)=3.85 \pm 0.12 \pm 0.40$ MeV, $m_{s}^{\msb}(2\ \gev)=105
\pm 3 \pm 9$ MeV and $m_s/m_{ud}=27.3 \pm 0.3 \pm 1.2$. We also obtain
$f_K=161.7 \pm 1.2 \pm 3.1$ MeV and the ratio $f_K/f_\pi=1.227 \pm 0.009  \pm
0.024$. From this ratio, by using the experimental determination of $\Gamma(K
\to \mu \bar \nu_\mu (\gamma))/\Gamma(\pi \to \mu \bar \nu_\mu (\gamma))$ and
the average value of $\vert V_{ud}\vert$ from nuclear beta decays, we obtain
$\vert V_{us}\vert=0.2192(5)(45)$, in agreement with the determination from
$K_{l3}$ decays and the unitarity constraint.}
\end{abstract}

\section{Introduction}
\label{sec:intro}

In this paper we extend to the kaon sector our previous lattice study of the
pion mass and decay constant~\cite{Boucaud:2007uk}. We present a determination
of the light quark masses, strange quark mass $m_s$ and the average up-down
quark mass $m_{ud}$, of the kaon pseudoscalar decay constant $f_K$, and of the
ratio $f_K/f_\pi$. We have simulated the theory with $N_f=2$ dynamical quarks,
taken to be degenerate in mass, and two valence quarks. In order to investigate
the properties of the $K$ meson, we consider in the present analysis a partially
quenched setup, namely we take the valence quark masses $\mu_1$ and $\mu_2$
different in value between each other and different from the sea quark mass
$\mu_S$.

The strategy of the calculation is the following. We first compute the
pseudoscalar meson masses and decay constants for different values of the sea
and valence quark masses, and study their mass dependence. We then use the
experimental values of the ratios $M_\pi/f_\pi$ and $M_K/M_\pi$ to determine the
average up-down and the strange quark mass respectively. The lattice spacing is
fixed from $f_\pi$. The results obtained for the quark masses are finally used
to evaluate $f_K$ and the ratio $f_K/f_\pi$.

The calculation is based on a set of gauge field configurations generated with
the tree-level improved Symanzik gauge action at $\beta=3.9$, corresponding to
$a=0.087(1)$ fm ($a^{-1} \simeq 2.3$ GeV)~\cite{Boucaud:2007uk}, and the
twisted mass fermionic action at maximal twist. We have simulated 5 values of
the bare sea quark mass,
\beq
a\mu_S=\{0.0040,\, 0.0064,\, 0.0085,\, 0.0100,\, 0.0150\}\, ,
\eeq
and computed quark propagators for 8 values of the valence quark mass,
\beq
a\mu_{1,2}=\{0.0040,\, 0.0064,\, 0.0085,\, 0.0100,\, 0.0150,\, 0.0220,\,
0.0270,\, 0.0320\}\ .
\eeq
The first five masses are equal to the sea quark masses, and lie in the range
$1/6\, m_s \simle \mu_{1,2} \simle 2/3\, m_s$, where $m_s$ is the physical
strange quark mass, while the heaviest three are around the strange quark mass.

We implement non-degenerate valence quarks in the twisted mass formulation of
lattice QCD as discussed for instance in
refs.~\cite{Frezzotti:2004wz,AbdelRehim:2006ve}. We introduce two twisted
doublets of degenerate valence quarks, ($u_1$, $d_1$) and ($u_2$, $d_2$), with
masses $\mu_1$ and $\mu_2$ respectively, and simulate charged mesons $\bar u_1
d_2$ and $\bar d_1 u_2$. Within each doublet, the two valence quarks are
regularized in the physical basis with Wilson parameters of opposite values
($r_u=-r_d=1$).

At each value of the sea quark mass we have computed the two-point correlation
functions of charged pseudoscalar mesons, with both degenerate and non
degenerate valence quarks, on a set of 240 independent gauge field
configurations, separated by 20 HMC trajectories one from the other (each
trajectory being of length 1/2). To improve the statistical accuracy, we have
evaluated the meson correlators using a stochastic method to include all spatial
sources. The method involves a real stochastic source ($Z(2)$-noise) for all
colour and spatial indices at one Euclidean time slice randomly moved when
passing from one gauge configuration to another. This ``one-end" method is
similar to that pioneered in ref.~\cite{Foster:1998vw} and implemented in
ref.~\cite{McNeile:2006bz}. Statistical errors on the meson masses and decay
constants are evaluated using the jackknife procedure, by decimating 10
configurations out of 240 in each jackknife bin. Statistical errors on the fit
results, which are based on data obtained at different sea quark masses, are
evaluated using a bootstrap procedure. Further details on the numerical
simulation can be found in~\cite{ETMC-long}.

The use of twisted mass fermions in the present calculation turns out to be
beneficial in several aspects~\cite{Frezzotti:2003ni,Frezzotti:2004wz}: i) the
pseudoscalar meson masses and decay constants, which represent the basic
ingredients of the calculation, are automatically improved at ${\cal
O}(a)$;~\footnote{Strictly speaking, automatic O(a) improvement was proved
in~\cite{Frezzotti:2003ni,Frezzotti:2004wz} to hold in a unitary as well as in
a mixed action framework. Actually the same proof goes through also in the
present partially quenched setup. The reason is that all the symmetries entering
the discussion of the renormalizability and O($a$) improvement are valid for
generic values of the masses of the various valence and sea quarks.} ii) once
maximal twist is realized, the physical quark mass is directly related to the
twisted mass parameter of the action, and it is subject only to multiplicative
renormalization; iii) the determination of the pseudoscalar decay constant does
not require the introduction of any renormalization constant, and it is based on
the relation
\beq
f_{PS} = (\mu_1 + \mu_2) \, \frac{\vert \langle 0 \vert P^1(0) \vert P
\rangle\vert} {M_{PS}^{\,2}}\, .
\eeq
Concerning the size of discretization effects, it is worth noting that, since
the two valence quarks are regularized in the physical basis with Wilson
parameters of opposite values, the meson mass $M_{PS}^{\,2}$ differs from its
continuum counterpart only by terms of ${\cal O}(a^2\mu)$ and ${\cal O}(a^4)$,
whereas $f_{PS}$ differs from its continuum limit by terms of ${\cal
O}(a^2)$~\cite{Sharpe:2004ny,Frezzotti:2005gi}. Therefore, at ${\cal O}(a^2)$
the cutoff effects on $M_{PS}^{\,2}$ and $f_{PS}$ are as in a chiral invariant
lattice formulation.

\begin{figure}[t]
\begin{center}
\includegraphics[scale=0.4,angle=270]{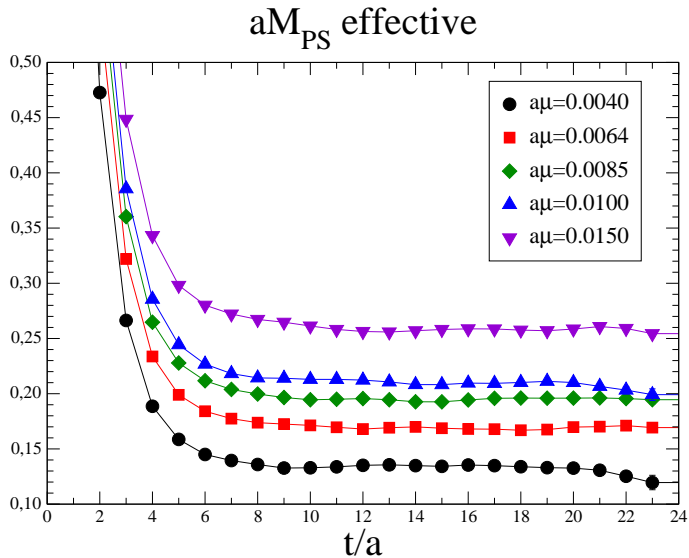}
\end{center}
\vspace{-1.0cm}
\caption{\small\sl Effective masses of pseudoscalar mesons, as a function of the
time, in the degenerate cases $\mu_S=\mu_1=\mu_2$. Error bars are smaller than
the symbol sizes.}
\label{fig:effmass}
\end{figure}

The meson mass $M_{PS}$ and the matrix element $\vert\langle 0 \vert P^1(0)
\vert P \rangle\vert$ have been extracted from a fit of the two-point
pseudoscalar correlation function in the time interval $t/a \in [10,21]$. In
order to illustrate the quality of the data, we show in fig.~\ref{fig:effmass}
the effective masses of pseudoscalar mesons, as a function of the time, in the
degenerate cases $\mu_S=\mu_1=\mu_2$.

\section{Quark mass dependence of pseudoscalar meson mas\-ses and decay
constants}
The determination of the physical properties of $K$ mesons requires a study of
the quark mass dependence of the corresponding observables over a large range of
masses, extending from the physical strange quark down to the light up and down
quarks. In this work, we study the quark mass dependence of the pseudoscalar
meson masses and decay constants by investigating two different functional
forms. The first one is the dependence predicted by continuum partially quenched
chiral perturbation theory (PQChPT), whereas in the second case we consider a
simple polynomial dependence. For a recent precision study of the quark mass
dependence of meson masses and decay constants in the partially quenched theory
with $N_f=2$ dynamical fermions see also ref.~\cite{Del Debbio:2007pz}.

\subsection{PQChPT fits}
Within PQChPT we consider the full next-to-leading order (NLO) expressions with
the addition of the local NNLO contributions, i.e. terms quadratic in the quark
masses, which turn out to be needed for a good description of the meson masses
and decay constants up to the region of quark masses around the strange quark
mass. The PQChPT predictions have been derived in ref.~\cite{sharpe} and can be
written in the form
\bea
\label{eq:mf12}
&& M_{PS}^{\,2}(\mu_S,\mu_1,\mu_2) \ =\ B_0 \left(\mu_1 + \mu_2\right) \cdot
\left[ 1 + \dfrac{\xi_1\, (\xi_S-\xi_1)\, \ln 2 \xi_1}{(\xi_2-\xi_1)} -
\dfrac{\xi_2\, (\xi_S-\xi_2)\, \ln 2 \xi_2}{(\xi_2-\xi_1)} + \nn \right.\\
&& \qquad
\left. + a_V\, \xi_{12} + a_S\, \xi_S + a_{VV}\, \xi_{12}^2 + a_{SS}\, \xi_S^2 
+ a_{VS}\, \xi_{12}\, \xi_S + a_{VD}\, \xi_{D12}^2 \right] \, , \\
&& f_{PS}(\mu_S,\mu_1,\mu_2) \ =\ f \cdot
\left[ 1 - \xi_{1S}\, \ln 2 \xi_{1S} - \xi_{2S}\, \ln 2 \xi_{2S} +
\dfrac{\xi_1\,\xi_2 - \xi_S\,\xi_{12}}{2\,(\xi_2-\xi_1)} \ln \left( 
\frac{\xi_1}{\xi_2}\right) + \nn \right.\\
&& \quad \left. + (b_V + 1/2) \, \xi_{12} + (b_S - 1/2)\, \xi_S + b_{VV}\,
\xi_{12}^2 + b_{SS}\, \xi_S^2 + b_{VS}\, \xi_{12}\, \xi_S + b_{VD}\, \xi_{D12}^2
\right]\, , \nn
\eea
where $\xi_i =2 B_0 \mu_i/(4\pi f)^2$, $\xi_{ij}=B_0 (\mu_i+\mu_j)/(4\pi f)^2$
and $\xi_{Dij}=B_0 (\mu_i-\mu_j)/(4\pi f)^2$. The parameters $B_0$ and $f$ are
the low energy constants (LECs) entering the chiral Lagrangian at the
LO~\footnote{The pseudoscalar decay constant $f$ is normalised such that $f_\pi
= 130.7$ MeV at the physical pion mass.}, whereas $a_V$, $a_S$, $b_V$ and $b_S$
are related to the NLO LECs~\cite{sharpe} by
\beq
\label{eq:lec}
a_V = 4 \alpha_8 - 2 \alpha_5 \quad , \quad a_S = 8 \alpha_6 - 4 \alpha_4 
\quad , \quad b_V = \alpha_5 \quad , \quad b_S = 2  \alpha_4 \, .
\eeq
The quadratic terms in the quark masses in eq.~(\ref{eq:mf12}) represent the
local NNLO contributions. The corresponding chiral logarithms at two loops in
the partially quenched theory are also known~\cite{Bijnens:2005pa}. They
involve, however, a larger number of NLO LECs whose values, in the $N_f=2$
theory, cannot be fixed from phenomenology. Introducing their contribution in
the fit would increase significantly the number of free parameters, thus
limiting, at the same time, the predictive power of the calculation.

In the limit of degenerate valence quark masses, $\mu_1=\mu_2 \equiv \mu_V$,
eq.~(\ref{eq:mf12}) is finite and reduces to
\bea
\label{eq:mfvv}
&& \hspace{-0.7cm}
M_{PS}^{\,2}(\mu_S,\mu_V,\mu_V) =  2\, B_0\, \mu_V \, \cdot \left[ 
1 + (2\,\xi_V-\xi_S)\, \ln 2 \xi_V + (a_V+1)\, \xi_V + (a_S-1)\, \xi_S +\right.
\nn \\
&& \hspace{-0.7cm}
\qquad \qquad \qquad \qquad \qquad \qquad \quad 
\left. + a_{VV} \,\xi_V^2 + a_{SS}\, \xi_S^2 + a_{VS}\, \xi_V\, \xi_S \right]
\, , \\
&& \hspace{-0.7cm}
f_{PS}(\mu_S,\mu_V,\mu_V) = f \cdot
\left[ 1 - 2\, \xi_{VS}\, \ln 2 \xi_{VS} + b_V \, \xi_V + b_S\, \xi_S  +
b_{VV}\, \xi_V^2 + b_{SS}\, \xi_S^2 + b_{VS}\, \xi_V\, \xi_S \right]
\,. \nn
\eea

\subsubsection{Finite volume corrections}
In a lattice QCD calculation aiming at a percent precision on the physical
predictions, the impact of finite size corrections cannot be neglected. The
lattice in our simulation has spatial extension $L=24 a \simeq 2.1$ fm, and the
pseudoscalar meson mass at the lightest value of the quark mass is such that
$M_{PS} L \simeq 3.2$. Since we have not performed yet a systematic study of
non-degenerate meson masses and decay constants on different lattice volumes, we
will estimate the finite size effects by including in the fits the corrections
predicted by one-loop chiral perturbation theory, which, in the partially
quenched case, are expressed by~\cite{bv}~\footnote{We thank D.Becirevic and
G.Villadoro for having provided us with the expression of finite volume
corrections to $M_{PS}^{\,2}(\mu_S,\mu_1,\mu_2)$ which is not given in
ref.~\cite{bv}.}
\bea
\label{eq:mf12fse}
&& \hspace{-0.7cm}
M_{PS}^{\,2}(\mu_S,\mu_1,\mu_2; L) = M_{PS}^{\,2}(\mu_S,\mu_1,\mu_2) \cdot 
\left[ 1 + \dfrac{\xi_1\, (\xi_S-\xi_1)\,\tilde g_1(L,\xi_1)}{(\xi_2-\xi_1)}
- \dfrac{\xi_2\, (\xi_S-\xi_2)\,\tilde g_1(L,\xi_2)}{(\xi_2-\xi_1)}\right]\,,
\nn \\
&& \hspace{-0.7cm}
f_{PS}(\mu_S,\mu_1,\mu_2; L) = f_{PS}(\mu_S,\mu_1,\mu_2) \cdot \nn \\
&&
\qquad \left[ 1 - \xi_{1S}\,\tilde g_1(L,\xi_{1S}) - \xi_{2S}\,\tilde
g_1(L,\xi_{2S})+ \dfrac{\xi_{12}-\xi_S}{2\,(\xi_2-\xi_1)} \, 
\left(\xi_1\,\tilde g_1(L,\xi_1) - \xi_2\,\tilde g_1(L,\xi_2)\right) +
\right. \nn \\
&&
\qquad \left. + \frac{1}{4}\, (\xi_{S}-\xi_1)\, \tilde g_2(L,\xi_1) +
\frac{1}{4}\, (\xi_{S}-\xi_2)\, \tilde g_2(L,\xi_2)\right]\,.
\eea
The functions $\tilde g_s$ ($s=1,2$) in eq.~(\ref{eq:mf12fse}) are defined as
\beq
\tilde g_s(L,M^2) = \dfrac{(4\pi)^{3/2}}{(M^2)^{2-s}} \, \Gamma(s-1/2) \,
\xi_{s-1/2}(L,M^2)\, ,
\eeq
where $M$ is the pseudoscalar meson mass at the LO, $M^2 = 2 B_0 \mu =(4\pi f)^2
\xi$,
\beq
\xi_s(L,M^2)=
\frac{1}{(4\pi)^{3/2}\Gamma(s)}\int_0^{\infty}d\tau\ \tau^{s-5/2}e^{-\tau M^2} 
\left[  \vartheta^3\left(\frac{L^2}{4\tau}\right)-1  \right] \, ,
\eeq
and $\vartheta(\tau)$ is the elliptic theta function
\beq
\vartheta(\tau) \equiv  \sum_{n=-\infty}^{\infty} e^{-\tau\, n^2} \,.
\eeq

The limits of eq.~(\ref{eq:mf12fse}) in the case of degenerate valence quark
masses, $\mu_1=\mu_2 \equiv \mu_V$, can be obtained by using the identity
\beq
M^2 \, \frac{d}{d M^2} \, \tilde g_s(L,M^2) = - (2-s)\, \tilde g_s(L,M^2) - 
\tilde g_{s+1}(L,M^2)
\eeq
and are given by
\bea
\label{eq:mvvfse}
&& M_{PS}^{\,2}(\mu_S,\mu_V,\mu_V; L) = M_{PS}^{\,2}(\mu_S,\mu_V,\mu_V) \cdot
\left[ 1 + \xi_V \, \tilde g_1(L,\xi_V) - (\xi_V-\xi_S) \, \tilde
g_2(L,\xi_V)\right]\,,
\nn \\
&& f_{PS}(\mu_S,\mu_V,\mu_V; L) = f_{PS}(\mu_S,\mu_V,\mu_V) \cdot 
\left[ 1 - 2\, \xi_{VS}\,\tilde g_1(L,\xi_{VS})\right]\,.
\eea

\subsection{Polynomial fits}
The inclusion of the local NNLO contributions in the PQChPT predictions
expressed by eq.~(\ref{eq:mf12}) is required by the observation that the pure
NLO predictions are not accurate enough to describe the quark mass dependence of
pseudoscalar meson masses and decay constants up to the region of the strange
quark. However, not having considered the full NNLO chiral predictions, we
regard eq.~(\ref{eq:mf12}) mostly as an effective description of the quark
mass dependence of these observables. In order to evaluate the associated
systematic uncertainty, we also consider in the analysis an alternative
description based on a simple polynomial dependence on the quark masses, for
both the pseudoscalar meson masses and decay constants:
\bea
\label{eq:mf12pol}
&& M_{PS}^{\,2}(\mu_S,\mu_1,\mu_2) \ =\ B_0\left(\mu_1 + \mu_2\right)\cdot\nn \\
&& \qquad  \qquad
\cdot \left[ 1 +  a_V\, \xi_{12} + a_S\, \xi_S + a_{VV}\, \xi_{12}^2 + a_{SS}\,
\xi_S^2 + a_{VS}\, \xi_{12}\, \xi_S + a_{VD}\, \xi_{D12}^2 \right] \, ,\\
&& f_{PS}(\mu_S,\mu_1,\mu_2) \ =\ f \cdot
\left[ 1  + (b_V + 1/2) \, \xi_{12} + (b_S - 1/2)\, \xi_S + b_{VV}\, \xi_{12}^2
+ b_{SS}\, \xi_S^2 + \nn \right.\\
&&  \qquad \qquad \left. + b_{VS}\, \xi_{12}\, \xi_S + b_{VD}\, \xi_{D12}^2
\right]\, . \nn
\eea
Note that, though we are adopting in eq.~(\ref{eq:mf12pol}) the same notation
for the coefficients of the chiral expansions as in eq.~(\ref{eq:mf12}), the
physical meaning of these coefficients, i.e. their relation to the derivatives
of $M_{PS}^{\,2}$ and $f_{PS}$ with respect to the quark masses, is actually
different. It also worth observing that, in the case of the polynomial
fits~(\ref{eq:mf12pol}), a change in the values of the LECs $f$ and $B_0$ only
amounts to a redefinition of the fit parameters of $M_{PS}^{\,2}$ and $f_{PS}$
respectively. Therefore, in this case, the two fits are independent one from the
other. The differences between the results obtained by performing either chiral
or polynomial fits will be included in the final estimates of the systematic
errors.

\section{Chiral extrapolations}
The input data in the present analysis are the lattice results for the
pseudoscalar meson masses and decay constants obtained at each value of the sea
quark mass, with both degenerate and non degenerate valence quarks. We exclude
from the fits the heaviest mesons having both the valence quark masses in the
strange mass region, namely with $a \mu_{1,2}=\{0.0220,\, 0.0270,\, 0.0320\}$.
Overall, we have considered therefore 150 combinations of quark masses for both
the meson masses and the decay constants. The full sets of results are collected
in tables~\ref{tab:Mps} and \ref{tab:Fps} of the appendix. The number of free
parameters in the combined fit of $M_{PS}^{\,2}$ and $f_{PS}$ is 14, but a first
analysis shows that some of them, in the various cases, are compatible with zero
within one standard deviation, and are kept fixed to zero in the final estimates
of the fit parameters (see table~\ref{tab:fitpar}).

In order to extrapolate the pseudoscalar meson masses and decay constants to the
points corresponding to the physical pion and kaon, we have considered
three different fits:
\begin{itemize}
\item {\bf\underline{Polynomial fit}:} a polynomial dependence on the quark
masses is assumed for the pseudoscalar meson masses and decay constants,
according to eq.~(\ref{eq:mf12pol}).
\item {\bf\underline{PQChPT fit}:} the pseudoscalar meson masses and decay
constants are fitted according to the predictions of PQChPT expressed by
eq.~(\ref{eq:mf12}) to which we add the finite volume corrections of
eq.~(\ref{eq:mf12fse}).
\item {\bf\underline{Constrained PQChPT fit}:} this fit, denoted as C-PQChPT in
the following, deserves a more detailed explanation. The main uncertainty in
using eqs.~(\ref{eq:mf12}) and (\ref{eq:mf12pol}) to effectively describe the
quark mass dependence of $M_{PS}^{\,2}$ and $f_{PS}$ is related to the
extrapolation toward the physical up and down quark masses. On the other hand,
we have shown in ref.~\cite{Boucaud:2007uk} that pure NLO ChPT, with the
inclusion of finite volume corrections, is sufficiently accurate in describing
the lattice results for both the pseudoscalar meson masses and decay constants
when the analysis is restricted to our lightest four quark masses in the unitary
setup (i.e. $\mu_1=\mu_2=\mu_S$). In order to take advantage of this
information, when performing the C-PQChPT fit we first determine the LO
parameters $B_0$ and $f$ and the NLO combinations $a_V+a_S$ and $b_V+b_S$ from a
fit based on pure NLO ChPT performed on the lightest four unitary points. In
other words, we repeat here as a preliminary step the same analysis done in
ref.~\cite{Boucaud:2007uk}, but on the smaller statistical sample of data used
for the present study.\footnote{Note that in the limit $\mu_1=\mu_2=\mu_S$, and
when all the coefficients of the quadratic terms are sent to zero, the PQChPT
expressions (\ref{eq:mf12}), as well as the finite volume corrections expressed
by eq.~(\ref{eq:mf12fse}), reduce to the pure NLO ChPT predictions used in the
chiral fit of ref.~\cite{Boucaud:2007uk}.} In this way we determine
\bea
\label{eq:unifit}
&& 2 a B_0 = 4.82(10) \ , \quad   af = 0.0552(12) \ , \nn \\
&& a_V+a_S = 0.80(23) \ , \quad   b_V+b_S = 0.62(24) \ .
\eea
These results, are perfectly consistent, at the level of $\sim 1.5 \ \sigma$,
with those obtained in ref.~\cite{Boucaud:2007uk}. By using the constraints of
eq.~(\ref{eq:unifit}), the other parameters entering the chiral expansions of
$M_{PS}^{\,2}$ and $f_{PS}$ are then obtained from a fit to eq.~(\ref{eq:mf12})
over the non unitary points. For consistency with the previous unitary fit, we
exclude also in this case from the analysis the data at the highest value of sea
quark mass, $a\mu_S=0.0150$.
\end{itemize}
In table~\ref{tab:fitpar} (``All data") we collect the results obtained for the
fit parameters in the three cases: polynomial, PQChPT and C-PQChPT fits. In the
last line we also quote the corresponding values of the $\chi^2$ per degree of
freedom. From these values we see that, though the quality of the fit is better
in the polynomial case, all three analyses provide a good description of the
lattice data, in the whole region of masses explored in the simulation. This is
only true, however, if the terms quadratic in the quark masses are taken into
account.
\begin{table}[!tb]
\begin{center}
\renewcommand{\arraystretch}{1.25}
\begin{tabular}{||c||c|c|c||c|c|c||} \cline{2-7} 
\multicolumn{1}{c||}{} & \multicolumn{3}{c||}{All data} &
\multicolumn{3}{c||}{Only $\mu_2 \ge \mu_1 = \mu_S$} \\  \hline
    Fit     & Polynomial &   PQChPT  & C-PQChPT  & Polynomial&   PQChPT   &
C-PQChPT  \\ \hline \hline
$ 2 a B_0 $ &
  4.59(3) &   4.79(6) &  4.82(10) &   4.55(6) &   4.86(12) &   4.82(10)\\
$   af    $ &
0.0607(6) & 0.0577(6) &0.0552(12) & 0.0606(9) & 0.0574(14) & 0.0552(12)\\ \hline
$  a_{V } $ &
 -0.63(7) &  2.37(10) &  2.15(18) & -0.52(16) &   1.91(15) &   2.15(18)\\
$  a_{S } $ &
    0.0   & -1.44(10) & -1.35(12) &     0.0   &  -1.04(37) &  -1.35(12)\\
$  b_{V } $ &
  2.66(4) &   0.68(5) &   0.86(8) &  2.56(13) &   0.49(12) &    0.75(8)\\
$  b_{S } $ &
 0.86(13) & -1.22(15) & -0.25(23) &  1.03(15) &  -0.94(34) &  -0.13(24)\\ \hline
$  a_{VV} $ &
   2.6(2) &   -9.3(3) &   -8.3(6) &    2.3(5) &   -7.8(18) &    -5.8(7)\\
$  a_{VS} $ &
    0.0   &    7.6(4) &    6.9(3) &     0.0   &    6.0(38) &     0.0   \\
$  a_{SS} $ &
    0.0   &     0.0   &     0.0   &     0.0   &     0.0    &     5.9(7)\\
$  a_{VD} $ &
  -0.6(1) &   -3.8(2) &   -3.2(3) &   -0.9(6) &   -2.6(21) &    -5.1(4)\\
$  b_{VV} $ &
  -4.0(2) &    1.2(2) &    0.9(1) &   -4.1(8) &     0.0    &     2.3(5)\\
$  b_{VS} $ &
    0.0   &    6.0(6) &   3.7(12) &     0.0   &    7.1(21) &     0.0   \\
$  b_{SS} $ &
    0.0   &     0.0   &  -5.3(14) &     0.0   &     0.0    &    -2.0(6)\\
$  b_{VD} $ &
  -3.7(2) &   -3.8(2) &   -3.0(3) &   -2.6(6) &     0.0    &    -3.1(6)\\ \hline
$\chi^2$/d.o.f. &
    0.38  &    1.34   &    1.11   &    0.28   &     0.40   &     0.78  \\
\hline
\end{tabular} 
\renewcommand{\arraystretch}{1.0}
\end{center}
\vspace{-0.4cm}
\caption{\sl Values of the fit parameters ~as obtained from the polynomial,
~PQChPT and C-PQChPT fits (see text for details), by analysing all combinations
of quark masses or only the combinations satisfying the constraint $\mu_2 \ge
\mu_1 = \mu_S$. In the last line, the corresponding $\chi^2$ per degree of
freedom are also given.}
\label{tab:fitpar}
\end{table}

A potential problem in the partially quenched theory is the divergence of the
chiral logarithms in the limit in which the light valence quark mass goes to
zero at fixed sea quark mass (see eq.~(\ref{eq:mf12})). This divergence does not
affect the extrapolation of the lattice results to the physical point, since the
sea and the light valence quark masses are degenerate in this case. However, in
order to verify that this unphysical behaviour of the partially quenched chiral
logarithms does not modify the result of the extrapolation, we have repeated the
analysis by restricting both the polynomial and the chiral fits to the $30$
quark mass combinations (26 in the case of the C-PQChPT fit) that, satisfying
the constraint $\mu_2 \ge \mu_1 = \mu_S$, are not affected by the dangerous
chiral logarithms. The results obtained for the free parameters of these fits
are also shown in table~\ref{tab:fitpar} (last three columns). By comparing
these results with those obtained by using the full set of data, we find some
differences in the estimates of the coefficients of the quadratic terms,
particularly those involving the sea quark mass ($a_{VS}$, $a_{SS}$, $\ldots$).
These differences reflect the relative importance in the fit of the various
quadratic terms in the different quark mass regions. For instance, in the case
of the highest sea quark mass, $a \mu_S=0.0150$, only 4 out of 30 combinations
of masses are included in the fit restricted by the condition $\mu_2 \ge \mu_1 =
\mu_S$. On the other hand, when we compare the results for the extrapolated
physical quantities ($am_{ud}$, $am_s$, $af_\pi$, $\ldots$) obtained from the
two fits, we find that they are almost indistinguishable (see
table~\ref{tab:results}). This is reassuring, as it shows that the effects of
potentially divergent chiral logarithms are well under control in our analysis.

\begin{figure}[p]
\begin{center}
\vspace{-2.4cm}
\includegraphics[scale=0.39,angle=270]{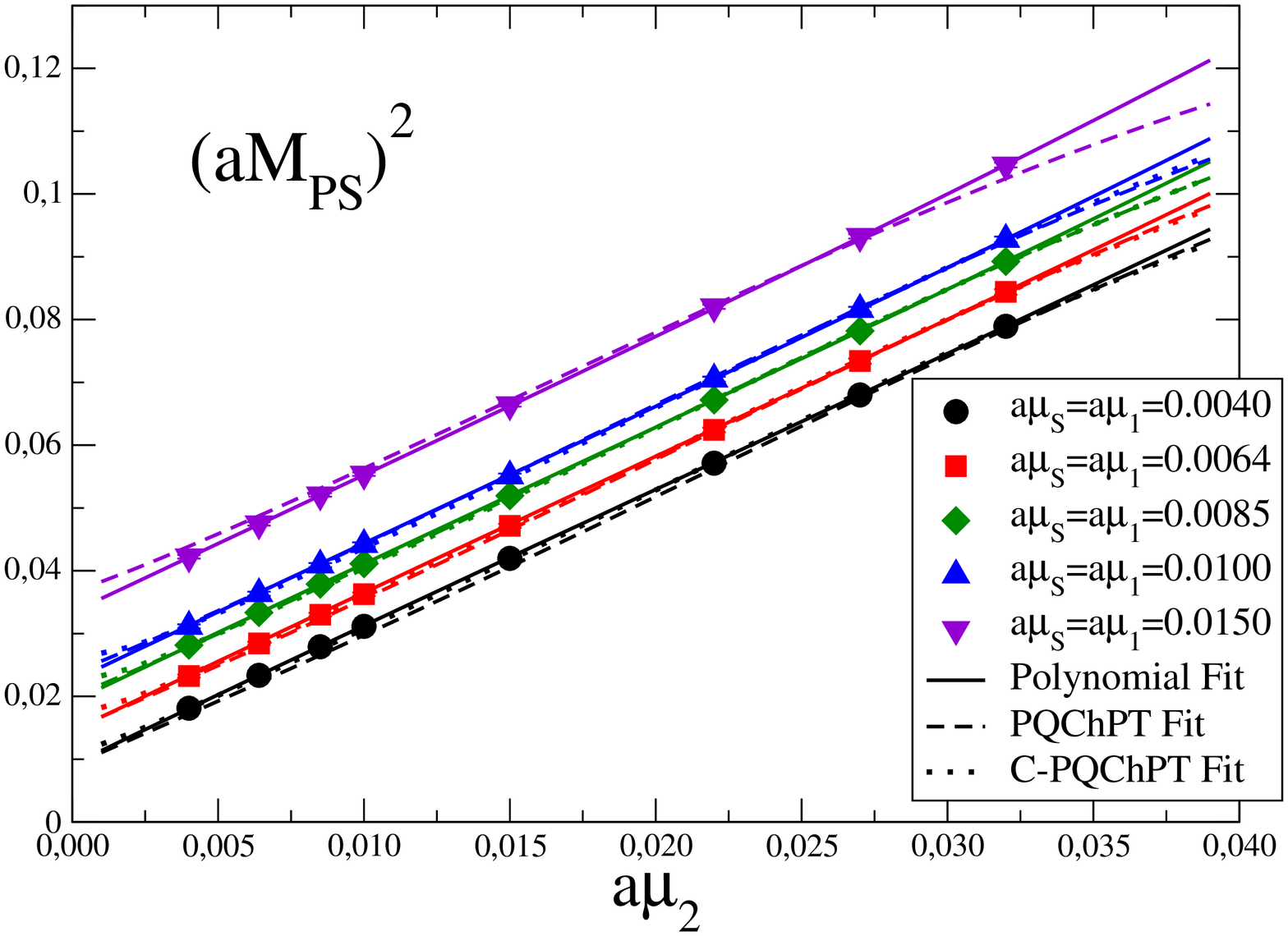} \\
\vspace{-0.5cm}
\includegraphics[scale=0.39,angle=270]{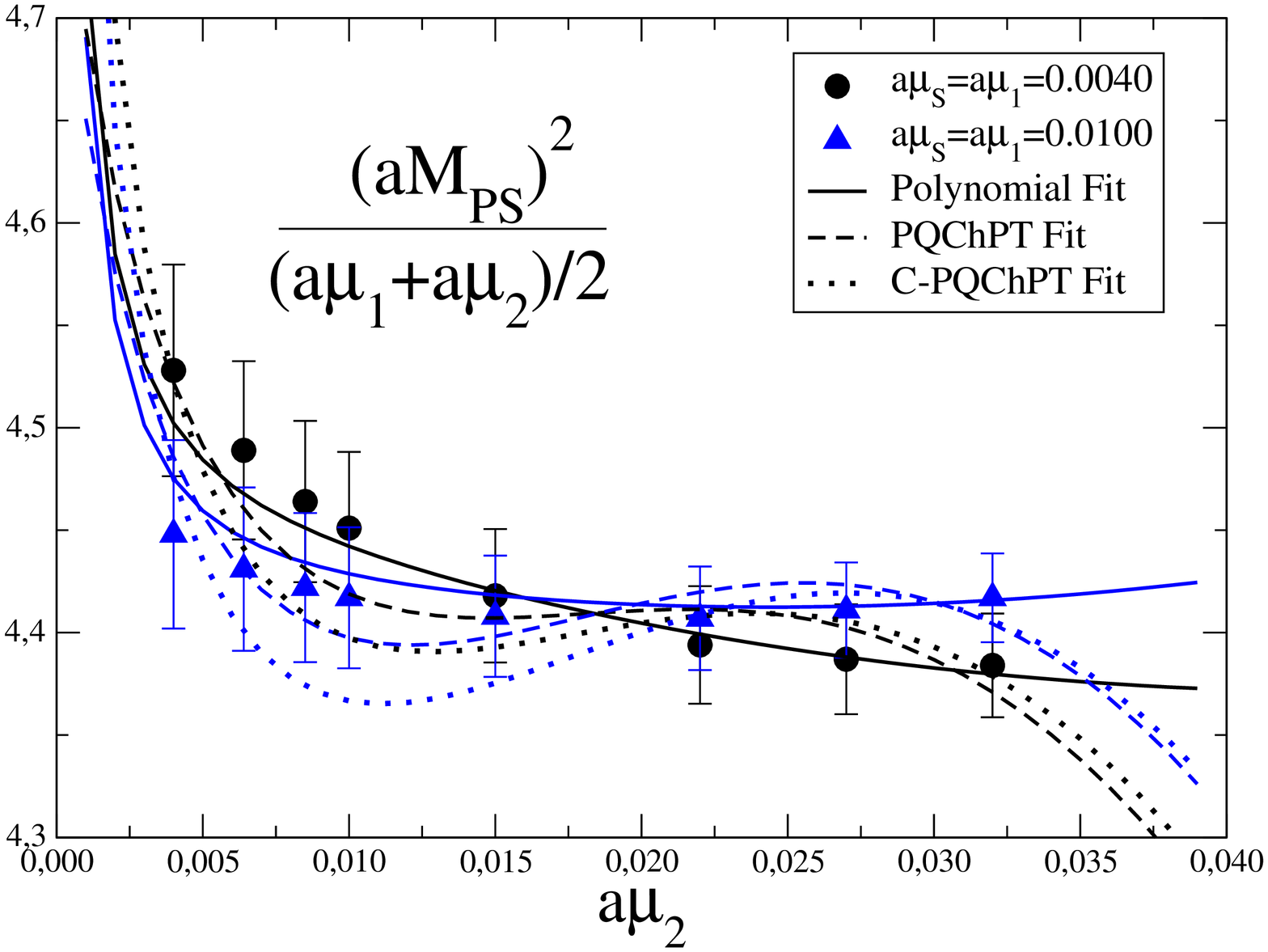} \\
\vspace{-0.5cm}
\includegraphics[scale=0.39,angle=270]{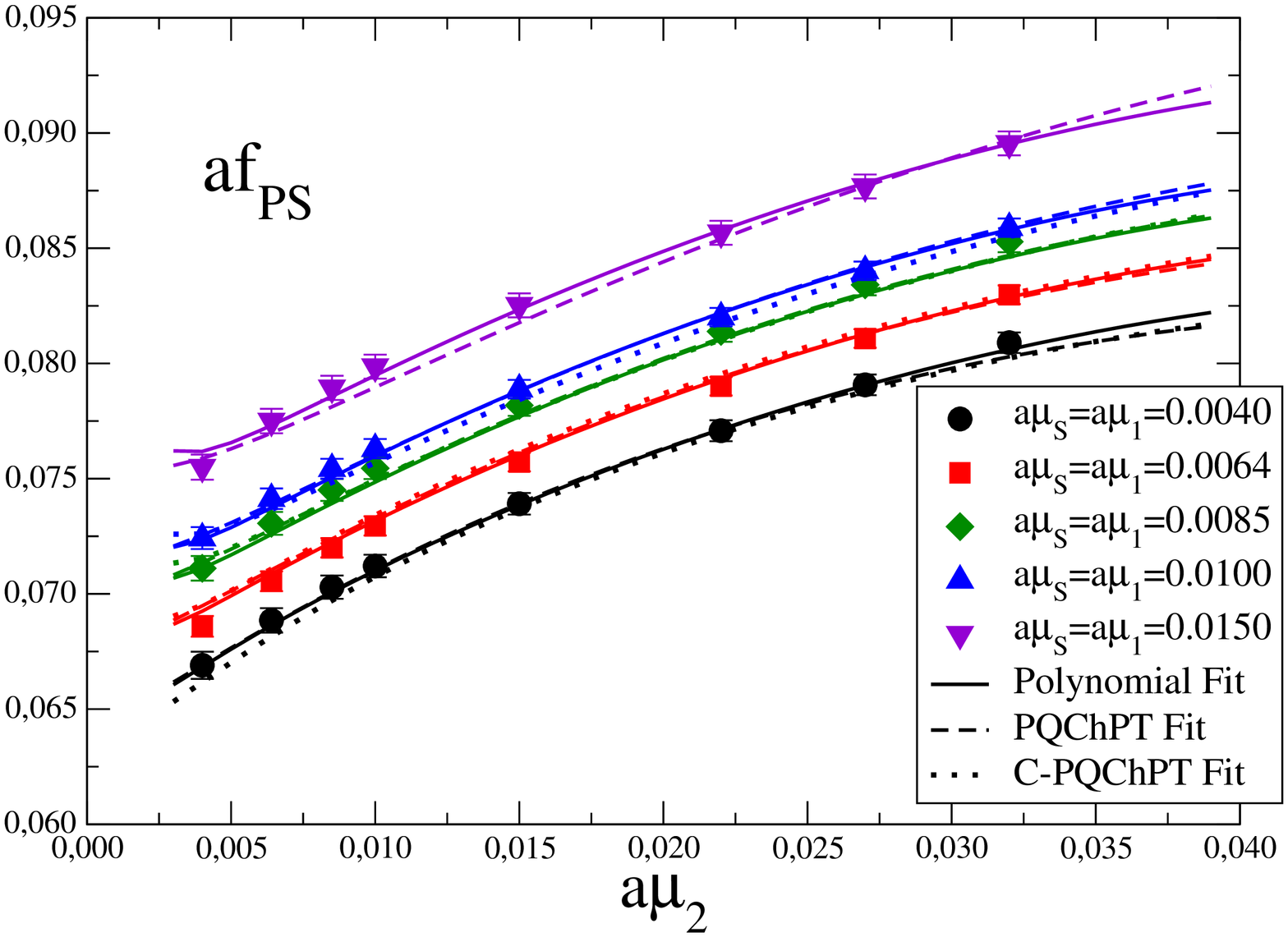}
\end{center}
\vspace{-1.2cm}
\caption{\sl Lattice results for $a^2 M_{PS}^{\,2}$ (top), $a^2 M_{PS}^{\,2}/
\frac{1}{2}(a\mu_1+a\mu_2)$ (center) and $a f_{PS}$ (bottom) as a function of
the valence quark mass $a\mu_2$, with $a\mu_1=a\mu_S$. The solid, dashed and
dotted curves represent the results of the polynomial, PQChPT and C-PQChPT fits
respectively. For better clarity, results at only two values of the see quark
mass have been shown in the center plot.}
\label{fig:3fits}
\end{figure}

The mass dependence of the pseudoscalar meson masses and decay constants is
illustrated in fig.~\ref{fig:3fits},
where we also compare the lattice data with the results of the polynomial,
PQChPT and C-PQChPT fits. We have shown in the plots the cases in which one of
the valence quark mass ($\mu_1$) is equal to the sea quark mass, and the results
are presented as a function of the second valence quark mass ($\mu_2$). The
points corresponding to the physical pion and kaon are thus obtained by
extrapolating/interpolating the results shown in fig.~\ref{fig:3fits} to the
limits $\mu_1 \to m_{ud}$ and $\mu_2 \to m_s$.

\begin{figure}[!t]
\begin{center}
\includegraphics[scale=0.4,angle=270]{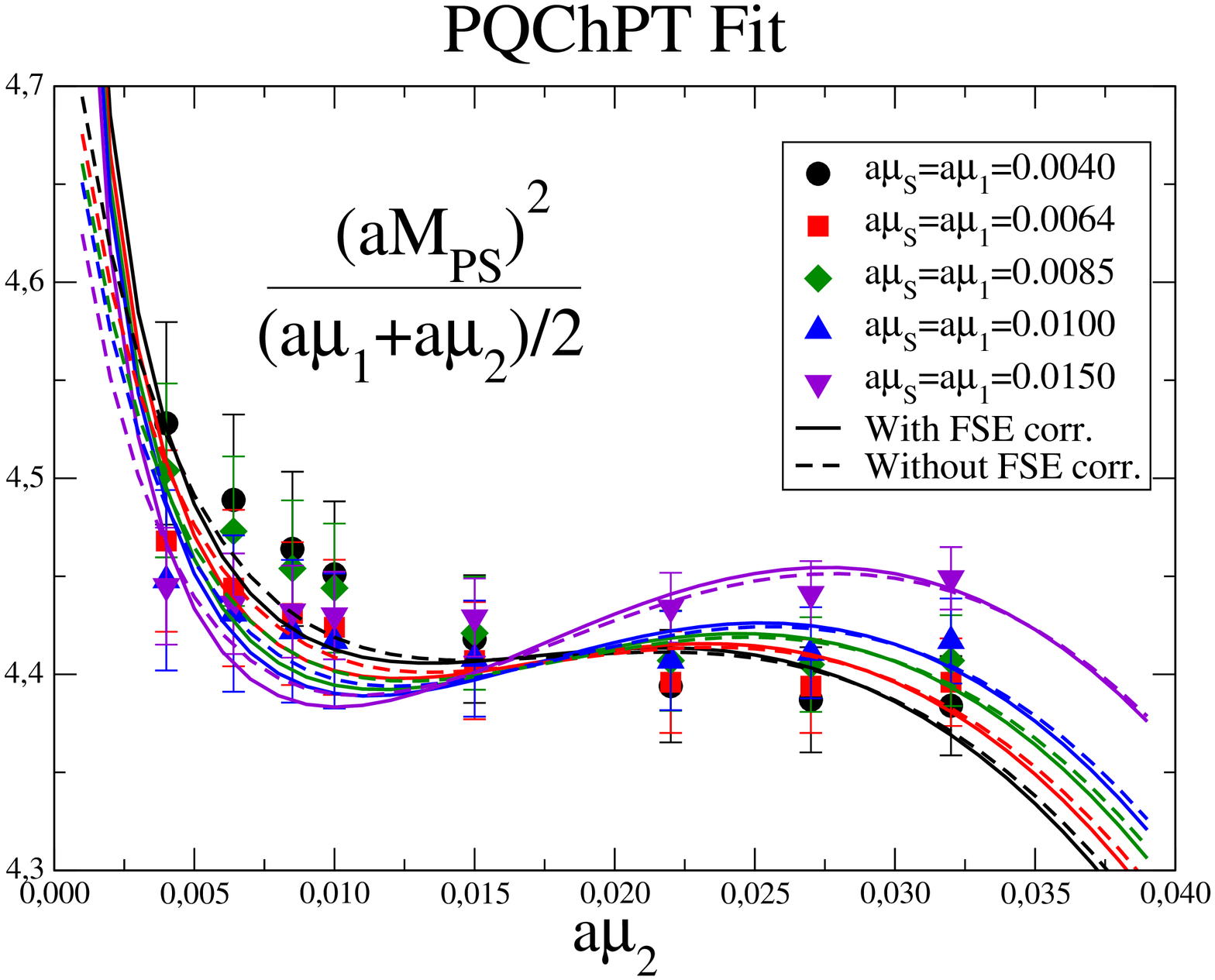}\\
\vspace{-0.5cm}
\includegraphics[scale=0.4,angle=270]{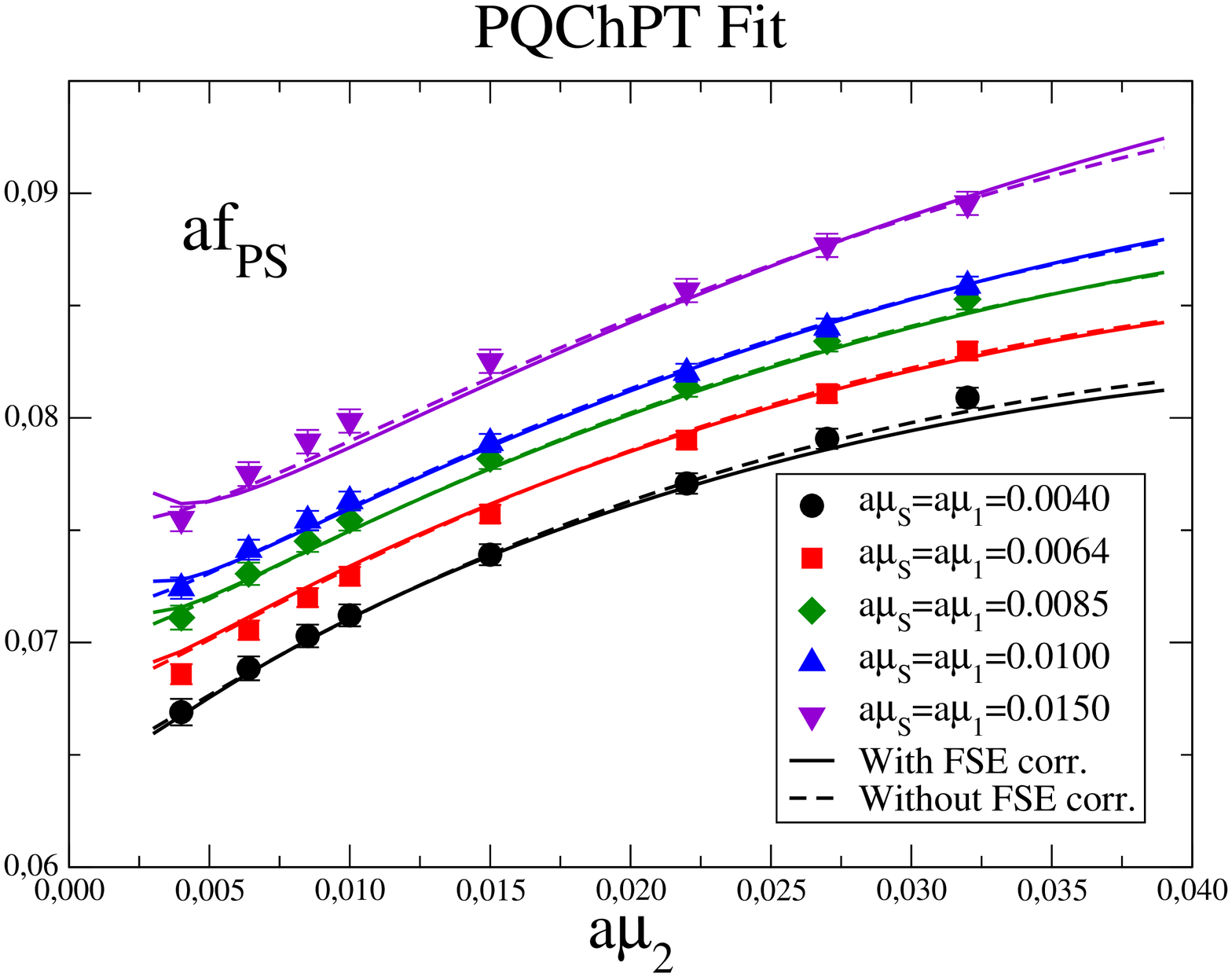}
\end{center}
\vspace{-1.0cm}
\caption{\small\sl PQChPT fits of the pseudoscalar meson masses and decay
constants performed with (solid lines) and without (dashed lines) including the
finite volume corrections of eq.~(\ref{eq:mf12fse}). The results are shown as a
function of the valence quark mass $a\mu_2$, with $a\mu_1 = a\mu_S$.}
\label{fig:fse}
\end{figure}

In order to illustrate the impact of finite volume corrections in the PQChPT
fits, we compare in fig.~\ref{fig:fse} the best fit curves for the pseudoscalar
meson masses and decay constants as obtained with or without including these
corrections. In the plots the differences between the two curves are barely
visible.
Obviously, a different question is whether the theoretical formulae based on
ChPT can accurately describe at the NLO the dependence of $M_{PS}^{\,2}$ and
$f_{PS}$ on the lattice volume. We postpone this issue to a future
investigation, in which we plan to better quantify the systematic error due to
finite size effects by extending the calculation of light pseudoscalar meson
masses and decay constants on lattices with different spatial sizes.

By having determined the fit parameters, we are now ready to extrapolate
eqs.~(\ref{eq:mf12}) and (\ref{eq:mf12pol}) to the physical pion and kaon. We
follow the procedure outlined in sect.~\ref{sec:intro}: we use the experimental
values of the ratios $M_\pi/f_\pi$ and $M_K/M_\pi$ to determine the average
up-down and the strange quark mass respectively. Once these masses have been
determined, we use again eqs.~(\ref{eq:mf12}) and (\ref{eq:mf12pol}) to compute
the values of the pion and kaon decay constants as well as their ratio
$f_K/f_\pi$.\footnote{In order to account for the electromagnetic isospin
breaking effects which are not introduced in the lattice simulation, we use as
``experimental" values of the pion and kaon mass the
combinations~\cite{Aubin:2004ck}
\[
(M_\pi^2)_{QCD} = M_{\pi^0}^2 \ , \quad (M_K^2)_{QCD} = \frac{1}{2} \left[
M_{K^0}^2 + M_{K^+}^2 - (1+\Delta_E) (M_{\pi^+}^2 - M_{\pi^0}^2) \right]
\]
with $\Delta_E=1$.}

In table~\ref{tab:results} we collect the values of the quark masses, meson
masses and decay constants, in lattice units, as obtained from the three fits by
analysing all combinations of quark masses or only the combinations that satisfy
the constraint $\mu_2 \ge \mu_1 = \mu_S$.
\begin{table}[!tb]
\begin{center}
\renewcommand{\arraystretch}{1.4}
\begin{tabular}{||c||c|c|c||c|c|c||} \cline{2-7} 
\multicolumn{1}{c||}{} & \multicolumn{3}{c||}{All data} &
\multicolumn{3}{c||}{Only $\mu_2 \ge \mu_1 = \mu_S$} \\  \hline
    Fit    & 
Polynomial & PQChPT & C-PQChPT & Polynomial & PQChPT & C-PQChPT \\ \hline \hline
$ a m_{ud}\cdot 10^3 $ &
   0.90(2) &   0.86(2) &   0.79(4) &   0.91(3) &    0.84(5) &    0.79(4) \\
$  a m_s  $      &
 0.0243(5) & 0.0235(5) & 0.0218(10)& 0.0243(7) & 0.0234(12) & 0.0217(10) \\
$  m_s/m_{ud}  $ &
   26.9(1) &   27.4(2) &   27.5(3) &   26.7(2) &    27.9(2) &    27.4(3) \\
$  a M_\pi     $ &
 0.0642(6) & 0.0632(6) & 0.0610(12)& 0.0642(9) & 0.0629(14) & 0.0610(12) \\
$  a M_K       $ &
  0.235(2) &  0.232(2) &  0.224(4) &  0.235(3) &   0.231(5) &   0.224(4) \\
$  a f_\pi     $ &
 0.0622(6) & 0.0612(6) & 0.0591(11)& 0.0622(8) & 0.0609(13) & 0.0591(11) \\
$  a f_K       $ &
 0.0756(7) & 0.0744(7) & 0.0730(11)& 0.0755(8) & 0.0747(11) & 0.0731(12) \\
$  f_K/f_\pi   $ &
  1.216(3) &  1.215(4) &  1.236(8) &  1.214(8) &  1.225(11) &   1.238(7) \\
\hline
\end{tabular}
\renewcommand{\arraystretch}{1.0}
\end{center}
\vspace{-0.4cm}
\caption{\sl Values of the quark masses, meson masses and decay constants in
lattice units as obtained from the polynomial, PQChPT and C-PQChPT fits by
analysing all combinations of quark masses or only the combinations satisfying
the constraint $\mu_2 \ge \mu_1 = \mu_S$.}
\label{tab:results}
\end{table}
Note that the values of the quark mass ratio $m_s/m_{ud}$ and of the ratio of
decay constants $f_K/f_\pi$, being dimensionless and well normalised quantities,
are obtained at this step without need of fixing the scale nor of introducing
the quark mass renormalization constant. For these quantities, therefore, the
results presented in table~\ref{tab:results} already represent physical
predictions of the calculation.

As a further investigation, we have studied how the results for the quark masses
and decay constants change when the analysis is performed only on mesons with
degenerate valence quarks. In this case, we find values of quark mass in good
agreement with those given in table~(\ref{tab:results}), whereas for $f_K$ and
$f_K/f_\pi$ we obtain results that are larger by about $5$\% than those quoted
in the table.  This reflects the fact that the mass difference between valence
quarks represents only a small effect in meson masses, while it turns out to be
relevant in decay constants, at the present level of accuracy, as shown by the
contribution of the $a_{VD}$ and $b_{VD}$ terms respectively in the simple
polynomial fits (see table~\ref{tab:fitpar}).

\section{Physical results}
In order to convert into physical units the results obtained for the strange
quark mass and the kaon decay constants we fix the scale within each analysis
(polynomial, PQChPT and C-PQChPT fits) by using $f_\pi$ as physical input. This
choice deserves some discussion. By looking at table~\ref{tab:results}, we see
that the value of the pion decay constant in lattice units as obtained from the
C-PQChPT fit is in agreement, at the level of 1.4 $\sigma$, with the result of
our previous study, $af_\pi=0.0576(7)$~\cite{Boucaud:2007uk}. Indeed, from the
present analysis we obtain the estimate $a=0.089(2)$ fm, to be compared with the
determination $a=0.087(1)$ fm of ref.~\cite{Boucaud:2007uk}. We also find that
the estimate of the lattice spacing obtained from the C-PQChPT analysis
coincides with the one derived from the pure NLO ChPT analysis performed over
the lightest four unitary points. This is expected, since as explained before
the NLO unitary fit over the four lightest quark masses is used as a constraint
in the C-PQChPT analysis, and the effect of the quadratic terms which are left
out in the first fit is negligible in the evaluation of $f_\pi$. We then
conclude that the difference between the determination $a=0.089(2)$ fm and the
one given in ref.~\cite{Boucaud:2007uk} is a purely statistical effect and, as
such, is properly accounted for by the quoted statistical errors. In the
analyses based on the PQChPT and polynomial fits, instead, we obtain the
estimates $a=0.092(2)$ fm and $a=0.094(1)$ fm respectively. In this case, the
differences with respect to the C-PQChPT determination, which are at the level
of 3\% and 6\% respectively, have a systematic origin related to the uncertainty
in the chiral extrapolation.

As mentioned before, rather than choosing a common estimate of the scale for the
polynomial, PQChPT and C-PQChPT analyses, we prefer to fix the scale by relying
on the determination of $af_\pi$ as obtained within each separate fit. This
choice has the important advantage that also the pion and kaon masses are fixed
in this way to their physical values within each fit, since the experimental
results for the ratios $M_\pi/f_\pi$ and $M_K/M_\pi$ have been used to determine
the light and strange quark masses. Therefore, the absolute normalization of the
fit functions describing the quark mass dependence of both the meson masses and
the decay constants is always correct, independently of the assumptions done on
the chiral behaviour. As a result, we find that the systematic differences among
the various determinations of $am_s$ and $af_K$ given in
table~\ref{tab:results}, which are at the level of 6\% and 2\% respectively,
reduce by approximately a factor of two when the results are converted in
physical units. Nevertheless, in the case of the polynomial and PQChPT fits we
conservatively add in the calculation of the dimensionful quantities a 6\% and
3\% of systematic error coming from the different estimates of the scale.

The determination of the physical strange and up-down quark masses also requires
implementing a renormalization procedure. The relation between the bare twisted
mass at maximal twist, $\mu_q$, and the renormalized quark mass, $m_q$, is given
by
\beq
m_q(\mu_R) = Z_m(g^2,a \mu_R)\, \mu_q(a) \, ,
\eeq
where $\mu_R$ is the renormalization scale, conventionally fixed to 2 GeV for
the light quarks. $Z_m$ is the inverse of the flavour non-singlet pseudoscalar
density renormalization constant, $Z_m=Z_P^{-1}$. We have used the ${\cal
O}(a)$-improved non-perturbative RI-MOM determination of $Z_P$, which gives
$Z_P^{\rm{RI-MOM}}(1/a) =0.39(1)(2)$ at $\beta=3.9$~\cite{rimom}, and converted
the result to the $\msb$ scheme at the scale $\mu_R=2$ GeV by using
renormalization group improved continuum perturbation theory at the
N$^3$LO~\cite{Chetyrkin:1999pq}.

In table~\ref{tab:physical} we collect the results for the light quark masses
and pseudoscalar decay constants, in physical units, as obtained from the
polynomial, PQChPT and C-PQChPT fits. For completeness, we also show in the
table the results for the ratios $m_s/m_{ud}$ and $f_K/f_\pi$ already given in
table~\ref{tab:results}.
\begin{table}[t]
\begin{center}
\renewcommand{\arraystretch}{1.4}
\begin{tabular}{||c||c|c|c||} \hline 
    Fit               &  Polynomial   &   PQChPT      & C-PQChPT \\ \hline\hline
$m_{ud}^{\msb}$ (MeV) & 4.07(9)(33)   & 3.82(15)(25)  & 3.74(13)(21) \\
$m_s^{\msb}$ (MeV)    &   109(2)(9)   &    107(3)(7)  &    102(3)(6) \\
$m_s/m_{ud}$          &  26.7(2)(0)   & 27.9(2)(0)    &   27.4(3)(0) \\
$f_K$ (MeV)           & 158.7(11)(89) & 160.2(15)(54) & 161.8(10)(0) \\
$f_K/f_\pi$           & 1.214(8)(0)   & 1.225(11)(0)  &  1.238(7)(0) \\
\hline
\end{tabular}
\renewcommand{\arraystretch}{1.0}
\end{center}
\vspace{-0.4cm}
\caption{\sl Results for the light quark masses and pseudoscalar decay
constants, in physical units, as obtained from the polynomial, PQChPT and
C-PQChPT fits respectively, by analysing only the combinations of quark masses
satisfying the constraint $\mu_2 \ge \mu_1 = \mu_S$. The quoted errors are
statistical (first) and systematic (second), the latter coming from the
uncertainties in the determination of the lattice scale and of the quark mass
renormalization constant.}
\label{tab:physical}
\end{table}
To be conservative, we only consider from now on the results obtained from the
analysis of the quark mass combinations satisfying the constraint $\mu_2 \ge
\mu_1 = \mu_S$ which, though being affected by larger statistical errors, are
safe from the effects of the potentially divergent chiral logarithms. In
table~\ref{tab:physical} we quote as a systematic error within each fit the
uncertainty associated with the determination of the lattice spacing and of the
quark mass renormalization constant.

In order to derive our final estimates for the quark masses and decay constants,
we perform a weighted average of the results of the three analyses presented in
table~\ref{tab:physical} and conservatively add the whole spread among these
results to the systematic uncertainty. In this way, we obtain as our final
estimates of the light quark masses the results
\beq
\label{eq:mlight}
m_{ud}^{\msb}(2\ \gev)=3.85 \pm 0.12 \pm 0.40 \ \mev  \quad , \quad
m_{s}^{\msb}(2\ \gev)=105 \pm 3 \pm 9 \ \mev \, ,
\eeq
and the ratio
\beq
m_s/m_{ud}=27.3 \pm 0.3 \pm 1.2 \, ,
\eeq
where the first error is statistical and the second systematic.

For the kaon decay constant and the ratio $f_K/f_\pi$ we obtain the accurate
determinations
\beq
\label{eq:fK}
f_K=161.7 \pm 1.2 \pm 3.1 \ \mev \quad, \quad 
f_K/f_\pi=1.227 \pm 0.009  \pm 0.024 \,.
\eeq

\begin{figure}[t]
\begin{center}
\includegraphics[scale=0.4,angle=270]{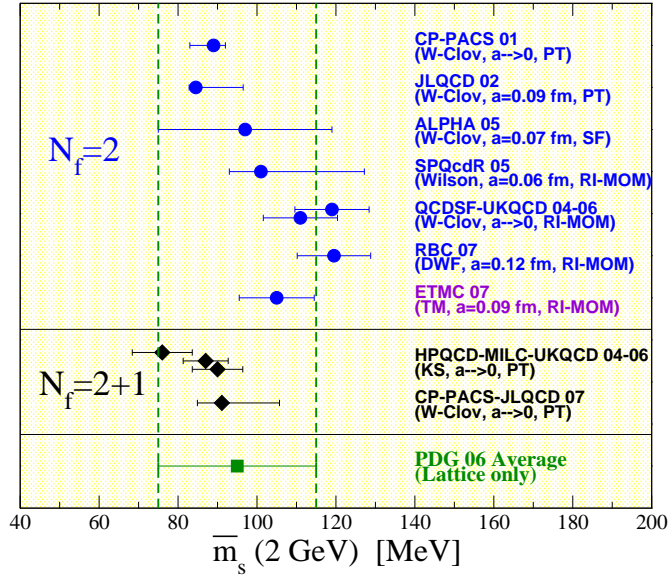}
\end{center}
\vspace{-1.0cm}
\caption{\small\sl Lattice QCD determinations of the strange quark mass obtained
from simulations with $N_f=2$~\cite{AliKhan:2001tx}-\cite{Blum:2007cy} and
$N_f=2+1$~\cite{Aubin:2004ck,Mason:2005bj,Bernard:2006wx,Ishikawa:2007nn}
dynamical fermions. The PDG average (from lattice only)~\cite{pdg} is also shown
for comparison.}
\label{fig:msunq}
\end{figure}

It is interesting to compare our result for the strange quark mass with other
lattice QCD determinations of the same quantity. This comparison is illustrated
in fig.~\ref{fig:msunq}.

The HPQCD-MILC-UKQCD Collaboration, using the MILC extensive simulations of
lattice QCD performed with $N_f=2+1$ dynamical improved staggered fermions,
initially quoted the result $m_s^{\msb}(2\ \gev)=76(3)(7)$
MeV~\cite{Aubin:2004ck}, significantly lower than our prediction in
eq.~(\ref{eq:mlight}). In~\cite{Aubin:2004ck}, the quark mass renormalization
constant was determined using one-loop perturbation theory. The two-loop
calculation has then led to a significant increase of the quark mass
estimate~\cite{Mason:2005bj}, and the most recent determination presented by
MILC now reads $m_s^{\msb}(2\ \gev)=90(5)(4)$ MeV~\cite{Bernard:2006wx}.
Recently, a similar result has been also obtained by the CP-PACS and JLQCD
Collaborations, using ${\cal O}(a)$-improved Wilson fermions with $N_f=2+1$ and
implementing the quark mass renormalization at one loop: $m_s^{\msb}(2\ \gev) =
91.1(^{+14.6}_{-6.2})$~\cite{Ishikawa:2007nn}. It is worth noting that this
result is perfectly consistent with the previous $N_f=2$ determinations of the
same quantity obtained by the two
collaborations~\cite{AliKhan:2001tx,Aoki:2002uc}.

In the present analysis, we find that the use of non-perturbative
renormalization plays a crucial role in the determination of the quark masses.
The estimate $Z_P^{\rm{RI-MOM}}(1/a)= 0.39(1)(2)$ obtained with the RI-MOM
method is in fact significantly smaller than the prediction $Z_P^{\rm{BPT}}(1/a)
\simeq 0.57(5)$ given by one-loop boosted perturbation theory (in the same
RI-MOM renormalization scheme)~\cite{rimom}. Had we used the perturbative
estimate of $Z_P$ we would have obtained $m_{ud}^{\msb}(2\ \gev)=2.63 \pm 0.08
\pm 0.36 \ \mev$ and $m_{s}^{\msb}(2\ \gev)=72 \pm 2 \pm 9 \ \mev$. As shown in
fig.~\ref{fig:msunq}, our prediction for the strange quark mass in
eq.~(\ref{eq:mlight}) is in good agreement with other determinations based on a
non-perturbative evaluation of the mass renormalization constant. These include
the results obtained by ALPHA, $m_s^{\msb}(2\ \gev) = 97(22)$
MeV~\cite{DellaMorte:2005kg}, by SPQ$_{\rm CD}$R, $m_s^{\msb}(2\ \gev) =
101(8)(^{+25}_{-0})$ MeV~\cite{Becirevic:2005ta}, by QCDSF-UKQCD, $m_s^{\msb}(2\
\gev) = 119(5)(8)$ MeV from the vector Ward identity~\cite{Gockeler:2004rp} and
$m_s^{\msb}(2\ \gev) = 111(6)(4)(6)$ MeV from the axial
one~\cite{Gockeler:2006jt}, and by RBC, $m_s^{\msb}(2\ \gev) = 119.5(56)(74)$
MeV~\cite{Blum:2007cy}. It is often found that, in lattice determinations of
quark masses, implementing a non-perturbative renormalization method has an
impact that can be larger even than the quenching effect. We believe that this
observation should be always kept in mind, particularly when the lattice results
for quark masses are considered for producing final averages.

\begin{figure}[t]
\begin{center}
\includegraphics[scale=0.4,angle=270]{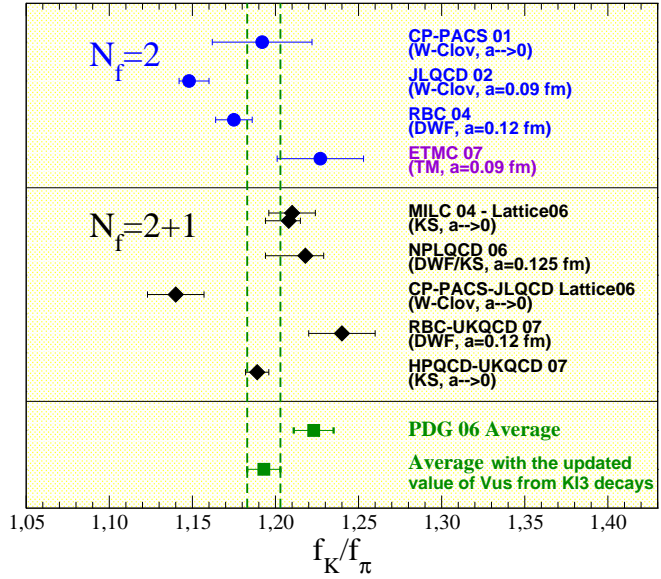}
\end{center}
\vspace{-1.0cm}
\caption{\small\sl Lattice QCD determinations of the ratio $f_K/f_\pi$ obtained
from simulations with $N_f=2$~\cite{AliKhan:2001tx,Aoki:2002uc,Aoki:2004ht} and
$N_f=2+1$~\cite{Bernard:2006wx},~\cite{Aubin:2004fs}-\cite{Follana:2007uv}
dynamical fermions. The results are also compared with the PDG 2006
average~\cite{pdg} and with the average based on the updated determination of
$V_{us}$ from $K_{\ell 3}$ decays~\cite{isidori}.}
\label{fig:fkfpiunq}
\end{figure}

Our result for the ratio $f_K/f_\pi$ is compared in fig.~\ref{fig:fkfpiunq} with
other recent lattice determinations based on simulations with $N_f=2$ and
$N_f=2+1$ dynamical fermions.

Our calculation and those based on the MILC improved staggered gauge
configurations are the only ones in which light quark masses significantly lower
than $m_s/3$ have been simulated ($m_q \simeq m_s/6$ for our lightest quark
mass). Therefore, it is interesting to compare our determination of $f_K/f_\pi$
with the more recent results quoted by MILC, $f_K/f_\pi=1.208(2)
(^{+7}_{-14})$~\cite{Bernard:2006wx}, and by HPQCD-UKQCD,
$f_K/f_\pi=1.189(7)$~\cite{Follana:2007uv}. Despite the strange quark is still
quenched in our simulation, and our results are still obtained at a single value
of the lattice spacing, we find the agreement between these determinations quite
satisfactory. In order to better quantify the size of discretization effects,
which are of ${\cal O}(a^2)$ in the present calculation, we plan to extend the
simulation to other two values of the lattice spacing (corresponding to
$\beta=3.8$ and $\beta=4.05$). This should also allow us to eventually perform
the extrapolation to the continuum limit.

Our result for the ratio $f_K/f_\pi$ can be combined with the experimental
measurement of $\Gamma(K \to \mu \bar \nu_\mu (\gamma))/\Gamma(\pi \to \mu \bar
\nu_\mu (\gamma))$~\cite{pdg} to get a determination of the ratio $\vert
V_{us}\vert/\vert V_{ud}\vert$~\cite{marciano}. We obtain
\beq
\label{eq:vuds}
\vert V_{us}\vert/\vert V_{ud}\vert= 0.2251(5)(47)\, ,
\eeq
where the first error is the experimental one and the second is the theory error
coming from the uncertainty on $f_K/f_\pi$. Eq~(\ref{eq:vuds}), combined with
the determination $\vert V_{ud} \vert= 0.97377(27)$~\cite{Marciano:2005ec} from
nuclear beta decays, yields the estimate
\beq
\vert V_{us}\vert= 0.2192(5)(45) \, ,
\eeq
in agreement with the value extracted from $K_{\ell 3}$ decays, $\vert V_{us}
\vert= 0.2255(19)$~\cite{isidori}, and leads to the constraint due to the
unitarity of the CKM matrix
\beq
\vert V_{ud}\vert^2 + \vert V_{us}\vert^2 + \vert V_{ub}\vert^2 -1 =
(-3.7 \pm 2.0)\cdot 10^{-3} \, .
\eeq

\section*{Acknowledgements}
We thank D.~Becirevic, G.~Martinelli and G.C.~Rossi for useful comments and
discussions.

The computer time for this project was made available to us by the John von
Neumann-Institute for Computing on the JUMP and  Jubl systems in J\"ulich and
apeNEXT system in Zeuthen, by UKQCD on the QCDOC machine at Edinburgh, by INFN
on the apeNEXT systems in Rome, by BSC on MareNostrum in Barcelona (www.bsc.es)
and by the Leibniz Computer centre in Munich on the Altix system. We thank
these computer centres and their staff for all technical advice and help.

This work has been supported in part by the DFG
Sonder\-for\-schungs\-be\-reich/Transregio SFB/TR9-03, the EU Integrated
Infrastructure Initiative Hadron Physics (I3HP) under contract
RII3-CT-2004-506078 and the EU contract MRTN-CT-2006-035482, ``FLAVIAnet". We
also thank the DEISA Consortium (co-funded by the EU, FP6 project 508830), for
support within the DEISA Extreme Computing Initiative (www.deisa.org).  R.F.,
V.L. and S.S. thank MIUR (Italy) for partial financial support under the
contracts PRIN06. V.G. and D.P. thank MEC (Spain) for partial financial
support under grant FPA2005-00711. M.P. acknowledges financial support by an EIF
Marie Curie fellowship of the European Community's Sixth Framework Programme
under contract number MEIF-CT-2006-040458.

\section*{Appendix}
In this appendix we collect in tables~\ref{tab:Mps} and \ref{tab:Fps} the values
of the pseudoscalar meson masses and decay constants obtained at the various
combinations of simulated sea and valence quark masses.

\begin{table}[p]
\vspace{-0.8cm}
\begin{center}
\renewcommand{\arraystretch}{1.5}
\begin{tabular}{||c|c||c|c|c|c|c||} \hline
$\mu_{1}$ & $\mu_{2}$ & $\mu_S=0.0040$ & $\mu_S=0.0064$ & $\mu_S=0.0085$ &
$\mu_S=0.0100$ & $\mu_S=0.0150$ \\ \hline
0.0040 & 0.0040 & 0.1346(8) & 0.1342(8) & 0.1357(10) & 0.1349(11) & 0.1364(9) \\
0.0040 & 0.0064 & 0.1528(7) & 0.1524(8) & 0.1537(9)  & 0.1529(10) & 0.1542(8) \\
0.0040 & 0.0085 & 0.1670(7) & 0.1667(8) & 0.1678(8)  & 0.1670(9)  & 0.1682(7) \\
0.0040 & 0.0100 & 0.1765(7) & 0.1762(8) & 0.1771(8)  & 0.1765(9)  & 0.1775(7) \\
0.0040 & 0.0150 & 0.2049(8) & 0.2046(8) & 0.2052(8)  & 0.2048(9)  & 0.2055(7) \\
0.0040 & 0.0220 & 0.2390(8) & 0.2388(8) & 0.2391(8)  & 0.2390(9)  & 0.2393(7) \\
0.0040 & 0.0270 & 0.2608(8) & 0.2606(8) & 0.2608(8)  & 0.2608(9)  & 0.2609(7) \\
0.0040 & 0.0320 & 0.2809(8) & 0.2808(8) & 0.2809(8)  & 0.2811(9)  & 0.2809(7) \\
0.0064 & 0.0064 & 0.1690(7) & 0.1687(8) & 0.1697(8)  & 0.1690(9)  & 0.1701(7) \\
0.0064 & 0.0085 & 0.1820(7) & 0.1817(7) & 0.1826(8)  & 0.1819(9)  & 0.1829(7) \\
0.0064 & 0.0100 & 0.1908(7) & 0.1905(7) & 0.1912(8)  & 0.1906(9)  & 0.1915(7) \\
0.0064 & 0.0150 & 0.2174(7) & 0.2171(7) & 0.2176(7)  & 0.2172(8)  & 0.2179(6) \\
0.0064 & 0.0220 & 0.2501(7) & 0.2499(7) & 0.2500(8)  & 0.2499(8)  & 0.2503(6) \\
0.0064 & 0.0270 & 0.2711(7) & 0.2709(7) & 0.2710(8)  & 0.2710(8)  & 0.2712(6) \\
0.0064 & 0.0320 & 0.2907(7) & 0.2905(7) & 0.2905(8)  & 0.2907(8)  & 0.2908(6) \\
0.0085 & 0.0085 & 0.1942(7) & 0.1939(7) & 0.1946(8)  & 0.1940(8)  & 0.1949(6) \\
0.0085 & 0.0100 & 0.2025(7) & 0.2022(7) & 0.2027(8)  & 0.2022(8)  & 0.2030(6) \\
0.0085 & 0.0150 & 0.2279(7) & 0.2276(7) & 0.2279(7)  & 0.2276(8)  & 0.2282(6) \\
0.0085 & 0.0220 & 0.2594(7) & 0.2592(7) & 0.2592(8)  & 0.2591(8)  & 0.2595(6) \\
0.0085 & 0.0270 & 0.2799(7) & 0.2796(7) & 0.2796(8)  & 0.2796(8)  & 0.2799(6) \\
0.0085 & 0.0320 & 0.2990(6) & 0.2988(7) & 0.2987(8)  & 0.2989(8)  & 0.2991(6) \\
0.0100 & 0.0100 & 0.2104(7) & 0.2101(7) & 0.2106(7)  & 0.2102(8)  & 0.2109(6) \\
0.0100 & 0.0150 & 0.2351(6) & 0.2348(7) & 0.2350(7)  & 0.2347(8)  & 0.2353(6) \\
0.0100 & 0.0220 & 0.2659(6) & 0.2656(7) & 0.2656(8)  & 0.2655(8)  & 0.2660(6) \\
0.0100 & 0.0270 & 0.2859(6) & 0.2857(7) & 0.2856(8)  & 0.2857(8)  & 0.2860(6) \\
0.0100 & 0.0320 & 0.3048(6) & 0.3046(7) & 0.3045(8)  & 0.3046(7)  & 0.3048(6) \\
0.0150 & 0.0150 & 0.2576(6) & 0.2573(7) & 0.2574(7)  & 0.2572(8)  & 0.2577(6) \\
0.0150 & 0.0220 & 0.2863(6) & 0.2861(7) & 0.2859(7)  & 0.2859(7)  & 0.2864(6) \\
0.0150 & 0.0270 & 0.3053(6) & 0.3050(7) & 0.3049(8)  & 0.3049(7)  & 0.3054(6) \\
0.0150 & 0.0320 & 0.3233(6) & 0.3230(7) & 0.3228(8)  & 0.3230(7)  & 0.3234(6) \\
\hline
\end{tabular}
\renewcommand{\arraystretch}{1.0}
\end{center}
\caption{\sl Values of the pseudoscalar meson masses
$aM_{PS}(\mu_S,\mu_1,\mu_2)$ for the various combinations of simulated sea and
valence quark masses.}
\label{tab:Mps}
\end{table}
\begin{table}[p]
\vspace{-0.8cm}
\begin{center}
\renewcommand{\arraystretch}{1.5}
\begin{tabular}{||c|c||c|c|c|c|c||} \hline
$\mu_{1}$ & $\mu_{2}$ & $\mu_S=0.0040$ & $\mu_S=0.0064$ & $\mu_S=0.0085$ &
$\mu_S=0.0100$ & $\mu_S=0.0150$ \\ \hline
0.0040 & 0.0040 & 0.0669(6) & 0.0666(5) & 0.0674(6) & 0.0681(6) & 0.0676(7) \\
0.0040 & 0.0064 & 0.0689(5) & 0.0686(5) & 0.0696(6) & 0.0701(5) & 0.0700(6) \\
0.0040 & 0.0085 & 0.0703(5) & 0.0701(4) & 0.0711(5) & 0.0715(5) & 0.0716(6) \\
0.0040 & 0.0100 & 0.0712(5) & 0.0710(4) & 0.0721(5) & 0.0724(5) & 0.0726(6) \\
0.0040 & 0.0150 & 0.0739(5) & 0.0738(4) & 0.0749(5) & 0.0751(4) & 0.0755(5) \\
0.0040 & 0.0220 & 0.0771(5) & 0.0772(4) & 0.0782(5) & 0.0783(4) & 0.0787(5) \\
0.0040 & 0.0270 & 0.0791(4) & 0.0792(4) & 0.0802(5) & 0.0804(4) & 0.0807(5) \\
0.0040 & 0.0320 & 0.0809(4) & 0.0811(4) & 0.0821(5) & 0.0822(4) & 0.0826(6) \\
0.0064 & 0.0064 & 0.0707(5) & 0.0706(4) & 0.0716(5) & 0.0719(5) & 0.0722(6) \\
0.0064 & 0.0085 & 0.0721(5) & 0.0720(4) & 0.0731(5) & 0.0732(5) & 0.0737(5) \\
0.0064 & 0.0100 & 0.0730(5) & 0.0729(4) & 0.0740(5) & 0.0741(4) & 0.0747(5) \\
0.0064 & 0.0150 & 0.0757(4) & 0.0757(4) & 0.0768(5) & 0.0768(4) & 0.0775(5) \\
0.0064 & 0.0220 & 0.0789(4) & 0.0790(4) & 0.0800(5) & 0.0799(4) & 0.0807(5) \\
0.0064 & 0.0270 & 0.0809(4) & 0.0811(4) & 0.0820(5) & 0.0819(4) & 0.0827(5) \\
0.0064 & 0.0320 & 0.0827(4) & 0.0830(4) & 0.0839(5) & 0.0838(4) & 0.0845(5) \\
0.0085 & 0.0085 & 0.0735(5) & 0.0734(4) & 0.0745(5) & 0.0746(4) & 0.0752(5) \\
0.0085 & 0.0100 & 0.0744(5) & 0.0744(4) & 0.0754(5) & 0.0754(4) & 0.0762(5) \\
0.0085 & 0.0150 & 0.0771(4) & 0.0771(4) & 0.0782(5) & 0.0780(4) & 0.0789(5) \\
0.0085 & 0.0220 & 0.0802(4) & 0.0804(4) & 0.0814(5) & 0.0812(4) & 0.0821(5) \\
0.0085 & 0.0270 & 0.0822(4) & 0.0825(4) & 0.0834(5) & 0.0832(4) & 0.0841(5) \\
0.0085 & 0.0320 & 0.0841(4) & 0.0844(4) & 0.0853(5) & 0.0850(4) & 0.0859(5) \\
0.0100 & 0.0100 & 0.0753(4) & 0.0753(4) & 0.0764(5) & 0.0763(4) & 0.0771(5) \\
0.0100 & 0.0150 & 0.0779(4) & 0.0780(4) & 0.0791(4) & 0.0789(4) & 0.0799(5) \\
0.0100 & 0.0220 & 0.0811(4) & 0.0813(4) & 0.0823(4) & 0.0820(4) & 0.0830(5) \\
0.0100 & 0.0270 & 0.0831(4) & 0.0834(4) & 0.0843(4) & 0.0840(4) & 0.0850(5) \\
0.0100 & 0.0320 & 0.0850(4) & 0.0853(4) & 0.0862(5) & 0.0859(4) & 0.0868(5) \\
0.0150 & 0.0150 & 0.0806(4) & 0.0808(4) & 0.0817(4) & 0.0814(4) & 0.0825(5) \\
0.0150 & 0.0220 & 0.0838(4) & 0.0841(4) & 0.0849(4) & 0.0846(4) & 0.0857(5) \\
0.0150 & 0.0270 & 0.0858(4) & 0.0862(4) & 0.0870(4) & 0.0866(4) & 0.0877(5) \\
0.0150 & 0.0320 & 0.0877(4) & 0.0881(4) & 0.0889(4) & 0.0885(4) & 0.0896(5) \\
\hline
\end{tabular}
\renewcommand{\arraystretch}{1.0}
\end{center}
\caption{\sl Values of the pseudoscalar decay constants $af_{PS}(\mu_S,\mu_1,
\mu_2)$ for the various combinations of simulated sea and valence quark masses.}
\label{tab:Fps}
\end{table}

\end{document}